\documentclass[a4paper]{article}

\usepackage{itgspeech2023}    
\usepackage{times}            
\usepackage[english]{babel}   
\usepackage[ansinew]{inputenc}
\usepackage[T1]{fontenc}      
\usepackage[sort&compress,numbers]{natbib}	
\usepackage{amsmath,amssymb}
\usepackage{graphicx}
\usepackage[colorlinks=false,pdfborder={0 0 0}]{hyperref}
\usepackage[acronym,toc,shortcuts,nonumberlist]{glossaries}
\usepackage{multirow}
\usepackage{pgfplots}
\usepackage{siunitx}
\usepackage{url}
\usepackage{xcolor}
\usepackage{xspace}
\usepackage{microtype}

\newacronym{AM}{AM}{acoustic model}
\newacronym{ASR}{ASR}{automatic speech recognition}
\newacronym{CTC}{CTC}{connectionist temporal classification}
\newacronym{DCT}{DCT}{discrete cosine transform}
\newacronym{FE}{FE}{feature extractor}
\newacronym{FIR}{FIR}{finite impulse response}
\newacronym{GPU}{GPU}{graphics processing unit}
\newacronym{LM}{LM}{language model}
\newacronym{NN}{NN}{neural network}
\newacronym{OCLR}{OCLR}{one-cycle learning rate}
\newacronym{SC}{SC}{supervised convolutional}
\newacronym{STFT}{STFT}{short time Fourier transform}
\newacronym{WER}{WER}{word error rate}

\newcommand\refSec[1]{Section~\ref{#1}}
\newcommand\refFig[1]{Figure~\ref{#1}}
\newcommand\refTab[1]{Table~\ref{#1}}
\newcommand{\ms}[1]{\SI{#1}{\milli\second}}
\newcommand{\devclean}{\textit{dev-clean}\xspace}
\newcommand{\devother}{\textit{dev-other}\xspace}
\newcommand{\testclean}{\textit{test-clean}\xspace}
\newcommand{\testother}{\textit{test-other}\xspace}
\newcommand{\relu}{ReLU\xspace}
\newcommand{\transformer}{\textit{Transformer}\xspace}
\newcommand{\conformer}{\textit{Conformer}\xspace}
\newcommand{\wv}{\textit{wav2vec}\xspace}
\newcommand{\wvtwo}{\textit{wav2vec 2.0}\xspace}  
\newcommand{\fe}{\gls{FE}\xspace}
\newcommand{\fes}{\glspl{FE}\xspace}

\title{Comparative Analysis of the wav2vec 2.0 Feature Extractor}

\author{Peter Vieting$^{1}$, Ralf Schl\"uter$^{1,2}$, Hermann Ney$^{1,2}$}

\address{%
  $^1$Machine Learning and Human Language Technology, RWTH Aachen University, Germany\\
  $^2$AppTek GmbH, Germany\\
  Email: \texttt{\{vieting,schlueter,ney\}@cs.rwth-aachen.de}}

\begin{document}

\maketitle

\begin{abstract}
\Gls{ASR} systems typically use handcrafted feature extraction pipelines.
To avoid their inherent information loss and to achieve more consistent modeling from speech to transcribed text, neural raw waveform \fes are an appealing approach.
Also the \wvtwo model, which has recently gained large popularity, uses a convolutional \fe which operates directly on the speech waveform.
However, it is not yet studied extensively in the literature.
In this work, we study its capability to replace the standard feature extraction methods in a \gls{CTC} \gls{ASR} model and compare it to an alternative neural \fe.
We show that both are competitive with traditional \fes on the LibriSpeech benchmark and analyze the effect of the individual components.
Furthermore, we analyze the learned filters and show that the most important information for the \gls{ASR} system is obtained by a set of bandpass filters.
\end{abstract}

\noindent\textbf{Index Terms}: speech recognition, feature extraction, raw waveform modeling, wav2vec 2.0

\glsresetall

\section{Introduction}
\Glspl{AM} for \gls{ASR} traditionally make use of hand-designed feature extraction methods like log Mel filterbank features or Gammatone features \cite{schlueter:icassp07}.
These techniques are motivated by insights into the properties of the human auditory system.
However, their inherent information loss can lead to suboptimal performance due to the fixed design.

This issue can be addressed by neural \fes.
Since their parameters are learned during training, these \fes can adapt to the need of \glspl{AM} and the loss of particularly helpful information may be avoided.
Several works have been proposed in this direction, often using convolutional \glspl{NN} \cite{palaz2015convolutional, golik15:cnn, tuske2018:waveform} but also applying parametrized functions \cite{ravanelli2018sincnet} and other architectures \cite{sainath2015cldnn}.

Recently, the \wvtwo framework \cite{facebook2020wav2vec2} has received large attention.
It is mainly known for its contrastive loss, which allows to pre-train a model on unlabeled audio data.
While the context network with multiple \transformer blocks constitutes the major part of the architecture, the \wvtwo model also uses a convolutional feature encoder.
This raises the question of how much this feature encoder contributes to the performance and whether it can be helpful for supervised \gls{ASR} as well.
Feature encoder \cite{facebook2020wav2vec2, facebook2020xlsr} and feature extractor \cite{asapp2022performance, vyas2022ondemand} are both common terms in the literature referring to the same part of the model and for consistency, we only use the term \acrfull{FE} here.

In this work, we show new insights into the \wvtwo \fe which is not yet closely studied in the literature.
Unlike previous work, we demonstrate that it can be used as a replacement for traditional feature extraction methods for a \gls{CTC} model similar to \gls{SC} features.
The results show that both are competitive on the LibriSpeech benchmark.
We analyze the learned filters and show the differences between the characteristics of both neural \fes.
Finally, we demonstrate that the learned bandpass filters of the \gls{SC} \fe encode the valuable information for the \gls{ASR} system while the remaining wideband filters are largely ignored.

\section{Related work}
This paper takes up the line of research that applies learnable feature extraction front-ends for \gls{ASR} \cite{palaz2015convolutional, golik15:cnn, tuske2018:waveform, ravanelli2018sincnet, sainath2015cldnn}.
In particular, different neural \fes are compared to traditional feature extraction methods.
We choose to investigate the supervised convolutional \fe from \cite{tuske2018:waveform}, which was also studied in \cite{vieting2021waveform}.
Furthermore, the feature extractor of \wvtwo \cite{facebook2020wav2vec2} is used as a learnable front-end for supervised \gls{ASR} training.

The \wv framework has been established in a series of related papers \cite{facebook2019wav2vec, facebook2019vqwav2vec, facebook2020wav2vec2, facebook2020xlsr}.
Besides modifications of the self-supervised training criteria -- moving from the future time step prediction \cite{facebook2019wav2vec} to masked time step prediction -- and incorporating quantization modules, also the architecture has been revised.
Instead of a fully convolutional architecture \cite{facebook2019wav2vec}, a large stack of \transformer blocks is introduced \cite{facebook2020wav2vec2}.
Also, the \fe itself has been adjusted,
where 5 layers, a group normalization with a single group and a \relu nonlinearity \cite{facebook2019wav2vec} are replaced by
7 layers with group normalization in the first layer and a GELU activation function \cite{facebook2020wav2vec2}.
Additionally, layer normalization and a linear projection are added after the convolutional layers.

While most works analyzing the \wvtwo model focus on the \transformer part which operates at a higher level of abstraction \cite{livescu2021wav2vec_analysis, fan21wvspeakerid, li2023exploration, dieck2022wav2vec}, some papers study the \fe as well.
The authors of \cite{choi2022w2v2fe} feed synthesized sine waves to a pre-trained model and observe that enough temporal detail is obtained and information about fundamental frequencies is represented in the model's output.
In particular, they claim the frequencies are distributed linearly over the Hertz scale, unlike perceptually motivated scales like the Mel scale or the Greenwood function for Gammatone features.
Also, the learned representations of the \fe are highly correlated with Mel spectrogram features, where the correlation is at a high plateau for \fe layers 3 to 7 \cite{livescu2021wav2vec_analysis}.
This observation is supported by further experimental evidence \cite{dieck2022wav2vec}.

Different settings for the \wvtwo \fe are studied in \cite{asapp2022performance}, showing that the required computation can be significantly reduced while maintaining similar performance.
In addition, a deeper variant with additional point-wise convolutional layers is proposed which outperforms its wider counterpart using higher inner dimensions.
However, unlike our work this is always done in the whole \wvtwo framework including pre-training and fine-tuning and no insights into the learned representations are given.

\section{Methods}
This work compares different \fes' applicability for \gls{ASR}.
We always normalize the waveform to zero mean and unit variance before it is input to an \fe.
The \fe is followed by the remaining \gls{AM}.
While the separation between \fe and remaining \gls{AM} is not sharp for neural feature extraction, we always refer to the part that replaces the traditional feature extraction as \fe.

\subsection{Feature Extractors}
\subsubsection{Standard Feature Extraction}
As a reference, we create baselines with standard, hand-designed \fes -- namely, log Mel filterbank features and Gammatone features.
To compute the log Mel filterbank features, first, the \gls{STFT} of the waveform is computed with a window size of \ms{25} and a shift of \ms{10}.
We keep the square of the magnitude and perform Mel warping to obtain an 80-dim vector.
Finally, $\log_{10}$ and normalization are applied.

Gammatone features apply a Gammatone filterbank to the pre-emphasized speech signal, perform temporal integration over each channel using a low pass filter, i.e., the Hanning window with a size of \ms{25} and a shift of \ms{10}, compress using the $10^{th}$ root and finally compute the \gls{DCT} of the values \cite{schlueter:icassp07}.
Further, the resulting 50-dim features are normalized.

\subsubsection{\wvtwo Feature Extractor}
The \gls{FE} of the \wvtwo model \cite{facebook2020wav2vec2} mainly consists of 7 convolutional layers.
They are configured with kernel sizes (10, 3, 3, 3, 3, 2, 2), strides (5, 2, 2, 2, 2, 2, 2), and a GELU activation function.
The first layer applies group normalization and finally, layer normalization and a linear projection are added.
The total receptive field of the \wvtwo \fe is \ms{25} and the shift between consecutive frames is \ms{20}.
Since the other \fes operate at a shift of \ms{10} and the \wvtwo \fe is built in a modular way, we remove the last layer with a stride of 2 to achieve features at the same frame rate.
This reduces the total receptive field to \ms{15}.

\subsubsection{Supervised Convolutional Features}
As an alternative comparable method for neural feature extraction, we use the \gls{SC} features \cite{tuske2018:waveform, vieting2021waveform}.
Similarly, a convolutional filterbank with learnable parameters is applied to the waveform.
As in the case of Gammatones, a temporal integration is performed over each channel.
However, in this case, multiple filters are used for temporal integration allowing for multi-resolutional processing.
Additionally, these filters are learned during training.
By default, we set the size of the first filterbank to 160, its stride to 10 and the number of channels to 150.
The second convolutional filter applies 5 temporal integration filters with a size of 40 and strides of 16 each.
Since the output of all 5 filters is stacked, we have a resulting feature dimension of 750.
The total receptive field of the \gls{SC} features is \ms{40}.

\subsection{Acoustic Model}
The \gls{AM} mainly consists of 3 VGG \cite{simonyan2015vgg} blocks for downsampling and 12 \conformer blocks.
Its configuration mostly follows \cite{zeineldeen2022conformer} and \cite{zhou2022efficient}.
The VGG downsampling uses three 3x3-convolutional layers with 32, 64 and 64 channels, respectively.
In total, a subsampling factor of 4 in the time dimension is achieved by the strided convolutions so that the \conformer operates on frames with a shift of \SI{40}{\milli\second}.
The feature dimension is reduced by a factor of 2 due to max pooling and then multiplied by the number of channels in the last convolution resulting in a total increase by factor 32 in our case.
After a linear projection to a 512-dimensional representation and dropout of 0.1, the 12 \conformer blocks with an inner dimension of 512, swapped convolution and multi-headed self-attention modules and 8 heads are applied.

\subsection{Connectionist Temporal Classification}
We use a \gls{CTC} model for our experiments.
It follows the setup described for the \gls{CTC} model in \cite{zhou2022efficient}.
All models are trained in a purely supervised fashion using the \gls{CTC} objective unless explicitly stated otherwise.
The official LibriSpeech 4-gram \gls{LM} is used and integrated using shallow fusion \cite{gulcehre2015shallow} during decoding.

\section{Experimental Setup}
\subsection{Data}
We conduct experiments on the 960h LibriSpeech corpus \cite{panayotov2015librispeech}, which consists of read English speech.
If the \wvtwo \fe is pre-trained (only in \refSec{sec:w2v_pretrain}), the same 960h were used in pre-training.
The \gls{LM} was trained on the default LibriSpeech \gls{LM} training data which includes the transcriptions of the training set along with the additional 800M-word text-only data.
The evaluation is done on the standard \devclean, \devother, \testclean and \testother sets.

\subsection{Training Details}
We generally follow the setup in \cite{zhou2022efficient}.
The batch size is 640k samples, which corresponds to \SI{40}{\second} of the speech waveform and the gradients are accumulated over 3 steps.
The NAdam optimizer uses a \gls{OCLR} schedule with a peak learning rate of $3\cdot 10^{-4}$ and we train the model for 20 epochs.
In contrast to \cite{zhou2022efficient}, no gradient noise is applied.
Each model is trained on a single \gls{GPU}.
The experiments are carried out using RETURNN \cite{zeyer2018returnn} and RASR \cite{rybach2011rasr}.

\section{Results}
The results, namely the \glspl{WER} on the LibriSpeech \textit{dev} and \textit{test} sets, for the different \fes are shown in \refTab{table:features_general}.
The baseline performance using Gammatone features is improved compared to \cite{zhou2022efficient} mainly by disabling gradient noise and re-tuning the peak learning rate.
We can observe that all \fes are within a close range.
The \gls{SC} features slightly outperform the Gammatone baseline, while log Mel features and the \wvtwo \fe are again better with advantages on \devother and \testother, respectively.
This demonstrates that learnable neural \fes are competitive with hand-designed methods in a purely supervised setup with 960h of training data.
Notably, this is achieved in \gls{CTC} training from scratch without any external information.
In contrast, hybrid model training in prior works uses fixed alignments which are created with the help of spectral features \cite{palaz2015convolutional, golik15:cnn, tuske2018:waveform, ravanelli2018sincnet, sainath2015cldnn, vieting2021waveform}.
However, unlike e.g. for audio classification \cite{zeghidour2021leaf}, a clear advantage of these techniques cannot be deduced from the experimental results.

The number of parameters for Gammatone and log Mel features refers to the \gls{FIR} and Mel filterbanks, respectively, that we use in our implementation even though they are not trainable.
The differences in the total number of parameters are influenced by the size of the \fe, but also of the linear layer before the \conformer which grows large as the feature dimension increases.

\begin{table}[htbp]

\vspace{-0.3cm}
\centering
\caption{Comparison of different feature extraction methods for a CTC model on LibriSpeech (cln denotes clean, oth denotes other).}
\label{table:features_general}
\begin{tabular}{|c|c|c|c|S|c|S|c|}
\hline
\multicolumn{2}{|c|}{Feature Extractor} & \multicolumn{2}{c|}{\#Params} & \multicolumn{4}{c|}{{WER [\%]}} \\\cline{5-8}
                 \multicolumn{2}{|c|}{} &         \multicolumn{2}{c|}{} &      \multicolumn{2}{c|}{{dev}} & \multicolumn{2}{c|}{{test}} \\\hline
                                   Type &           Name &                         Total &   FE &                           {cln} & oth &                       {cln} & oth \\\hline\hline
                                  Fixed &      Gammatone &                         73.7M &  32k &                             3.0 & 7.1 &                         3.4 & 7.7 \\\cline{2-8}
                                        & Mel Filterbank &                         74.2M &  21k &                             2.9 & 6.9 &                         3.3 & 7.2 \\\hline
                                     NN &             SC &                         85.2M &  26k &                             2.9 & 7.0 &                         3.4 & 7.6 \\\cline{2-8}
                                        &    wav2vec 2.0 &                         89.5M & 4.1M &                             2.9 & 6.8 &                         3.3 & 7.5 \\
\hline
\end{tabular}

\vspace{-0.35cm}
\end{table}

\subsection{Dissecting the wav2vec 2.0 Feature Extractor}
\label{sec:w2v_components}
It is striking that the \wvtwo \fe uses two orders of magnitude more parameters than the \gls{SC} architecture.
To understand where \wvtwo uses them and how much they contribute to the \gls{FE}'s performance, we run it with different configurations.
The results for different widths and depths are shown in \refTab{table:features_w2v_size}.
The kernel sizes and strides were chosen as (10, 3, 3, 3, 3, 2), (5, 2, 2, 2, 2, 2) for 6 layers, (10, 6, 3, 3, 3), (5, 4, 2, 2, 2) for 5 layers, (10, 6, 6, 3), (5, 4, 4, 2) for 4 layers, (20, 6, 6), (10, 4, 4) for 3 layers and (32, 20), (16, 10) for 2 layers.
The results show that decreasing the inner dimension deteriorates the performance, however, a larger dimension does not improve over the 512-dim baseline.
In contrast, no major effect can be observed regarding the number of layers.
We also tried a variant with only one layer using kernel size 320 and stride 160 which did not converge.
Removing the final projection with about 394k parameters has no significant impact.

\begin{table}[htbp]

\vspace{-0.5cm}
\centering
\caption{Studying the effect of the wav2vec 2.0 feature extractor's width and depth.}
\label{table:features_w2v_size}
\begin{tabular}{|c|c|c|S|c|S|c|}
\hline
\#Layers &      Dim & \#Params FE & \multicolumn{4}{c|}{{WER [\%]}} \\\cline{4-7}
         &          &             &      \multicolumn{2}{c|}{{dev}} & \multicolumn{2}{c|}{{test}} \\\cline{4-7}
         &          &             &                         {clean} & other &                     {clean} & other \\\hline\hline
       6 &     1024 &       15.5M &                             2.9 &   6.8 &                         3.3 &   7.5 \\\cline{2-7}
         &      512 &        4.1M &                             2.9 &   6.8 &                         3.3 &   7.5 \\\cline{2-7}
         &      256 &        1.1M &                             2.9 &   7.1 &                         3.4 &   7.6 \\\cline{2-7}
         &      128 &        330k &                             3.0 &   7.2 &                         3.4 &   7.7 \\\cline{2-7}
         &       64 &        108k &                             3.1 &   7.4 &                         3.5 &   7.8 \\\hline
       5 &      512 &        4.3M &                             2.9 &   7.1 &                         3.4 &   7.6 \\\cline{2-7}
         &       64 &        112k &                             3.0 &   7.4 &                         3.5 &   7.9 \\\hline
       4 &      512 &        4.3M &                             2.9 &   7.0 &                         3.3 &   7.6 \\\cline{2-7}
         &       64 &        112k &                             3.1 &   7.2 &                         3.4 &   7.8 \\\hline
       3 &      512 &        3.6M &                             2.9 &   6.9 &                         3.3 &   7.4 \\\cline{2-7}
         &       64 &        101k &                             3.1 &   7.4 &                         3.5 &   7.9 \\\hline
       2 &      512 &        5.6M &                             3.0 &   7.1 &                         3.5 &   7.7 \\\cline{2-7}
         &       64 &        134k &                             3.1 &   7.6 &                         3.4 &   8.2 \\\hline
       6 &   64-512 &        1.0M &                             3.0 &   7.1 &                         3.4 &   7.6 \\\cline{2-7}
         & 128-1024 &        3.3M &                             2.9 &   7.0 &                         3.3 &   7.5 \\\cline{1-1}\cline{3-7}
      11 &          &        5.0M &                             2.9 &   6.9 &                         3.3 &   7.4 \\
\hline
\end{tabular}

\end{table}

The proposed \fe in \cite{asapp2022performance} uses increasing inner dimensions and we use this approach in the last three lines of \refTab{table:features_w2v_size}.
The first layer uses a dimension of 64/128 and it is doubled after each layer with an odd index.
Additionally, we add point-wise convolutional layers after all but the first layer using the same inner dimension as the preceding layer for the last row of the table.
It can be observed that these configurations perform similarly to the 6-layer 512-dim baseline while reducing the training time by about 21\%, 19\% and 14\%, respectively.
Note that the \gls{SC} features perform similarly to the variant with dimension 64-512 with again around 10\% shorter training time.
\begin{figure*}[htb]
	\centering
	\input{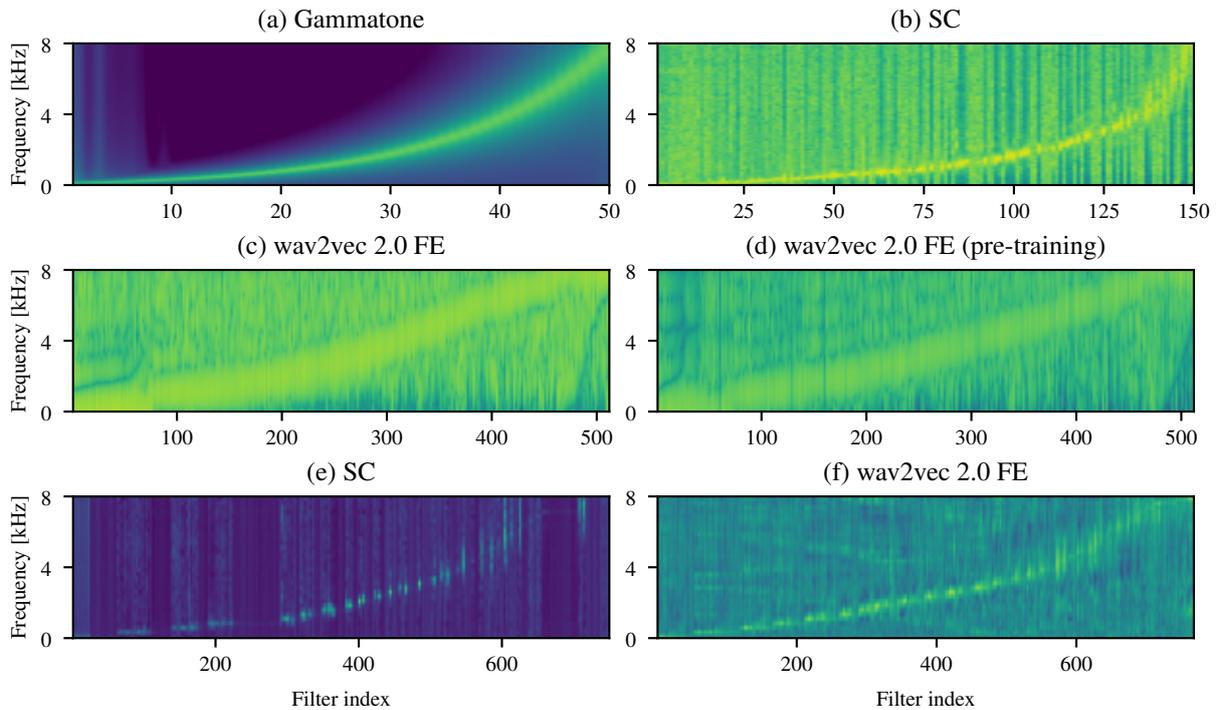}
	\vspace{-0.35cm}
	\caption{(a)-(d) show the frequency response of different \fes' filters in the first layer that operates on the waveform. (e)-(f) show the complete learned \fe network's response to single synthesized sine waves.\vspace{-0.35cm}}
	\label{fig:first_layer}
\end{figure*}

\subsection{\wvtwo Pre-Training}
\label{sec:w2v_pretrain}
The main appeal of the \wvtwo framework is the ability to pre-train the model on unlabeled audio data.
However, \cite{facebook2020wav2vec2} even reports relative \gls{WER} improvements of about 10\% on the \textit{other} subsets for the experiments on LibriSpeech when pre-training on the same data as used in supervised training compared to purely supervised training from scratch.
Here, we study this effect only for the \fe by using the parameters of an \fe pre-trained on LibriSpeech\footnote{\raggedright\url{https://github.com/facebookresearch/fairseq/blob/main/examples/wav2vec/README.md}}, so no external data is added.
During the supervised training, the \fe weights can be further trained or kept frozen.
As shown in \refTab{table:features_pretraining}, no positive effect can be observed here.
In fact, using a frozen pre-trained \fe shows clearly worse performance than training the \fe purely supervised from scratch.
We hypothesize that the pre-training mainly benefits the \transformer encoder and not the \fe while the mismatch between the \transformer encoder in pre-training and \conformer encoder in our setup could also pose a challenge for the frozen \fe.

\begin{table}[htbp]

\vspace{-0.3cm}
\centering
\caption{Effect of pre-training the wav2vec 2.0 \fe using the unsupervised loss on the same data (LibriSpeech 960h).}
\label{table:features_pretraining}
\begin{tabular}{|c|c|S|c|S|c|}
\hline
\multicolumn{2}{|c|}{Feature Extractor} & \multicolumn{4}{c|}{{WER [\%]}} \\\cline{3-6}
                 \multicolumn{2}{|c|}{} &      \multicolumn{2}{c|}{{dev}} & \multicolumn{2}{c|}{{test}} \\\hline
                              Trainable & Pre-trained &                         {clean} & other &                     {clean} & other \\\hline\hline
                                    yes &          no &                             2.9 &   6.8 &                         3.3 &   7.5 \\\cline{2-6}
                                        &         yes &                             2.9 &   6.9 &                         3.3 &   7.5 \\\cline{1-1}\cline{3-6}
                                     no &             &                             3.0 &   7.1 &                         3.3 &   7.8 \\
\hline
\end{tabular}

\vspace{-0.5cm}
\end{table}

\subsection{Frequency Response of Learned Filters}
\label{sec:freq_resp}
To analyze the learned filters of the neural \fes, we first plot the frequency response of the filters in the first layer which operates on the raw waveform.
As the order of learned filters in a convolutional layer is arbitrary, we sort the filters by the peak value of the frequency response and by the upper and lower \SI{3}{\decibel} cutoff frequencies as second and third criteria.
The plots are depicted in \refFig{fig:first_layer}\,(a)-(d), which share the same color range.

The well-known distribution of the Gammatone filters is given in \refFig{fig:first_layer}\,(a) as a reference.
\refFig{fig:first_layer}\,(b) shows the frequency response of the learned \gls{SC} filters.
In line with \cite{tuske2018:waveform}, bandpass filters with a non-linear distribution of center frequencies are learned.
This confirms, that these filters are also preferred by more advanced neural back-ends like the self-attention-based \conformer which are superior in the exploitation of context information compared to simpler neural back-ends e.g. in \cite{tuske2014raw}.
The distribution resembles the Gammatone filterbank and with higher center frequencies, the passband is also wider.
However, the stopband attenuation is clearly weaker and there are also several filters with no clear passband and a rather uncharacteristic frequency response.

The distribution for the \wvtwo \fe's first layer in \refFig{fig:first_layer}\,(c) does not entail this degree of resemblance.
Again, bandpass filters are learned, however, they have a wider passband, their center frequencies are distributed rather linearly and the stopband attenuation is weak.
In addition, about 14\% of all filters have their peak at \SI{0}{\kilo\hertz} and another 10\% at \SI{8}{\kilo\hertz} with varying bandwidths.
Several filters exist with multiple passbands.
Interestingly, the frequency response of the pre-trained \wvtwo \fe in \refFig{fig:first_layer}\,(d) is very similar to \refFig{fig:first_layer}\,(c).
This indicates that the properties learned in unsupervised pre-training and supervised \gls{CTC} training are similar.

\refFig{fig:first_layer}\,(e) and (f) show the complete \fe network's response to single synthesized sine waves of different frequencies, similar to the approach in \cite{choi2022w2v2fe}.
Unlike suggested in \cite{choi2022w2v2fe}, the distribution of \wvtwo center frequencies is not as linear anymore.
Moreover, the ratio of peak to mean values is much lower than for the \gls{SC} features.

\subsection{Filter Masking}
\begin{figure}[ht]
	\vspace{-0.25cm}
	\centering
	\input{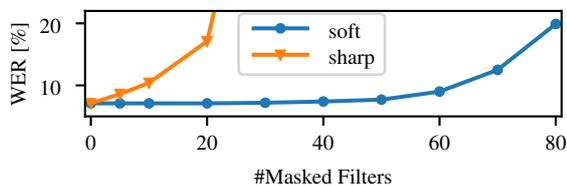}
	\vspace{-0.3cm}
	\caption{\gls{WER} [\%] on \devother using the \gls{SC} features, where $N$ out of 150 filters of the first layer are masked during recognition based on the peak-to-average ratio of their frequency response.}
	\label{fig:masked_filters}
\end{figure}

The observed uncharacteristic wideband filters from \refSec{sec:freq_resp} raise the question of whether their output is particularly helpful for the model in addition to the bandpass filters or if they are just not trained properly and ignored by the system.
To answer this question, we masked a portion of the 150 filters in the \gls{SC} features' first layer.
The filter kernels are sorted by the peak-to-average ratio of their frequency response and the $N$ filters with the highest (sharp) and lowest (soft) ratio are replaced by zeros during recognition.
Since there is no bias, the masked filters have a constant output of 0.
In \refFig{fig:masked_filters}, we can observe that masking the first 20 soft filters has no effect on the \gls{WER} (absolute difference <0.1\%) and only after masking 50, so a third of all filters, the \gls{WER} degrades significantly (7.7\%).
For the sharp filters, a clear impairment can be observed already when masking only 5 filters.
This demonstrates that the wideband filters, unlike the sharp ones, do not contain valuable information for the \gls{ASR} system and are rather the result of a suboptimal training process.
Future work should address this observation to allow better exploitation of the \fe's potential.

\section{Conclusions}
In this work, we utilize the \wvtwo \fe as a replacement for traditional feature extraction methods in a \gls{CTC} \gls{ASR} model on LibriSpeech.
It shows competitive performance and outperforms an alternative neural \fe, namely the \gls{SC} features, slightly.
We demonstrate that its width is more responsible for its capabilities than its depth and show that pre-training the \fe only is not beneficial.
Finally, we analyze the learned filters and highlight the differences between both neural \fes.
In particular, it is shown that the \gls{SC} \fe's learned bandpass filters encode the valuable information for the \gls{ASR} system while the remaining wideband filter information is largely ignored.

\section{Acknowledgements}
The authors thank Wei Zhou for providing the baseline \gls{CTC} model and Jan Rosendahl for helpful discussions.
This work was partially supported by the project HYKIST funded by the German Federal Ministry of Health on the basis of a decision of the German Federal Parliament (Bundestag) under funding ID ZMVI1-2520DAT04A.

\bibliographystyle{ieeetr}
\bibliography{mybib}

\end{document}